# Difficulties in the Implementation of Quantum Computers

Abhilash Ponnath

**Abstract:** This paper reviews various engineering hurdles facing the field of quantum computing. Specifically, problems related to decoherence, state preparation, error correction, and implementability of gates are considered.

## 1. Introduction

Faster speed and the capacity to solve hitherto unsolvable problems are the twin motivations driving research on the building of quantum computers. The use of information theory in quantum systems goes back more than thirty years [1-7]. It had two objectives: characterization of information in a quantum system [4,6], and using a quantum system to simulate a useful computation [7,8].

Current approaches to quantum computation exploit the phenomena of entanglement and superposition to create a paradigm that is more powerful than that of classical computing. The field has grown from the exotic arena confined to a few theoretical physicists into a full scale theoretical and experimental research area with millions of dollars being spent to build prototypes of quantum computers. Many believe that these computers are the only way that one can transcend the eventual limitations of Moore's law, where currently advances have been obtained mainly by advances in lithography.

In a quantum computer, the initial state of the quantum register together with the unitary transformations that need to be applied on this state depend of the problem. But this is done mathematically, and it is not clear whether the many hurdles that confront the actual building of these computers will ever be overcome. The main mathematical results relate to the exponential speed up provided by the quantum version of the Fourier transform, which was used by Shor [9] in the development of an fast method of factorization, and an algorithm by Grover [10] that searches an unordered list quadratically faster than any classical algorithm.

Quantum theoretic ideas have also been put to use in cryptographic applications [11-13], and although some of these cryptographic systems have been shown to be realizable, it appears that their cost –and the availability of equally effective classical schemes—will preclude their commercial adoption.

The requirements for physical implementation of a quantum computer include a well defined extendible qubit array for stable memory, feasible state preparation for the initial state, long decoherence time, a universal set of gate operations and capability for single-quantum measurement**s** [14,15]. Not all these requirements are currently unconditionally achievable. In Section 2, the sensitivity of the quantum hardware to interaction with the environment is discussed, and Section 3 deals with the matrix transformations required for quantum computation. Section 4 is a discussion of inevitable errors and proposals for error correction. Section 5 describes constraints on state preparation, and Section 6 discusses quantum information, uncertainty and issues of entropy of quantum gates.

**2 Sensitivity to interaction with environment**

Quantum computers are extremely sensitive to interaction with the surroundings since any interaction (or measurement) leads to a collapse of the state function. This phenomenon is called decoherence. It is extremely difficult to isolate a quantum system, especially an engineered one for a computation, without it getting entangled with the environment [16-18]. The larger the number of qubits the harder is it to maintain the coherence.

In a recent review [19], H.D. Zeh argues that decoherence cannot be reversed by redundancy coding: "[D]islocation of superpositions requires a distortion of the environment _by the system φ rather than a distortion of the system by the environment (such as by classical "noise"). This leads to the important consequence that decoherence in quantum computers cannot be error-corrected for in the usual manner by means of redundant information storage. Adding extra physical quantum bits to achieve redundancy, as it would be appropriate to correct spin or phase flips in the system, would in general even raise the quantum computer's vulnerability against decoherence – for the same reason as the increased size of an object normally strengthens its classicality. (Error correction codes proposed in the literature for this purpose are based on the presumption of decoherence-free auxiliary qubits, which may not be very realistic.)"

In another paper [20], the influence of spontaneous symmetry breaking on the decoherence of a many-particle quantum system was studied and it was shown that this symmetry breaking imposes a fundamental limit to the time that a system can stay quantum coherent. Miniaturization of the systems seems to lower the time limit. This means that the decoherence constraints on quantum computing are even more severe than thought before.

**3. Reliable matrix transformations**

Quantum computation on qubits is accomplished by operating upon them with an array of transformations that are implemented in principle using small gates [21,22]. It is imperative that no phase errors be introduced in these transformations. But practical schemes are likely to introduce such errors.

It is also possible that the quantum register is already entangled with the environment even before the beginning of the computation. Furthermore, uncertainty in initial phase makes calibration by rotation operation inadequate [23,24].

In addition, one must consider the relative lack of precision in the classical control that implements the matrix transformations. This lack of precision cannot be completely compensated for by the quantum algorithm.

**4 Errors and their correction**

Computation invariably involves errors, which are internal or externally induced. The classical computers has the capacity to perform well because the non-linearly (clamping or hard-limiting) of the computation process makes it possible to eliminate small errors, subsequent to which the larger bit errors can be eliminated using error-correction coding.

In the proposed quantum error correction schemes, the assumption is that only qubit is in error but the method would not work if we have more than one qubits error. Current error correcting algorithms [25-30] consider only bit flips, phase flips or both between the relative phases but these are not all the errors that we might encounter in a quantum computation.

Unlike the situation in classical computing, small errors cannot be eliminated in quantum computing [31] as it is a linear process, and it rules out operations analogous to clamping and hard-limiting. Furthermore, the no-cloning theorem prevents us from copying an unknown state and doing additional testing on it.

Random unitary transformation errors can occur in initializing a qubit [32]. The important characteristics of errors that need to be considered are component proliferation, nonlocal effects and amplitude errors. In a quantum register the errors can be due to a variety of reasons and we cannot group the errors in a systematic way. Errors perturbation of information in quantum arena is nonlocal compared to the local in classical arena and hence the concept of controlling errors using higher dimensional code word space is not realistic.

**5 Constraints on state preparation**

State preparation is the essential first step to be considered before the beginning of any quantum computation. In most schemes, the qubits need to be in a superposition state for the quantum computation to proceed correctly.

We have a variety of problems due to the nature of superposition and entanglements, and state transition using local transformations is not realistic in a large system. Macro systems that have been used as model quantum computing systems [14, 33,34] appear to implement not pure states but mixtures. Thus it appears that the NMR experiments do not validate quantum algorithm.

Quantum statistical constraints also need to be considered while designing algorithms to be used in quantum computers as they will directly affect the computation [35,36]. Since this is not always done, it appears that many models are unrealistic [37-39].

## 6 Quantum information, uncertainty and entropy of quantum gates

Classical information is easy to obtain by means of interaction with the system. On the other hand, the impossibility of cloning means that any specific unknown state cannot be determined. This means that unless the system has specifically been prepared, our ability to control it remains limited. The average information of a system is given by its entropy. The determination of entropy would depend on the statistics obeyed by the object [40,41].

If the gate is a physical device then from an information point of view its control can be characterized in terms of entropy. For example, consider the quantum teleportation protocol implemented with a rotation Hadamard gate [42]. The state cannot be recovered if the angle of rotation is less than the precision available on the receiving side. Computation with noisy components would require that the quantum circuit employed have the entropy rate smaller than the information capacity of the controller used.

Kak has argued [43]: "The control of the gate – a physical device – is by modifying some classical variable, which is subject to error. Since one cannot assume infinite precision in any control system, the implications of varying accuracy amongst different gates becomes an important problem. …[I]n certain arrangements a stuck fault cannot be reversed down the circuit stream using a single qubit operator, for it converted a pure state into a mixed state."

## 7 Conclusions

We have examined several problems associated with the implementation of quantum computers and we conclude that we currently do not possess technical solutions related to the fundamental tasks of quantum gate design, state preparation, and error correction.

The basic problem of implementation is a result of quantum computation being essentially an analog process. Using unitary transformations to solve real problems involving rotations, as in factorization, is an attractive mathematical idea, but there remain fundamental engineering hurdles in the implementation of this idea.